\pdfoutput=1

\documentclass[twocolumn,amsmath,amssymb]{revtex4}
\usepackage{graphicx}

\begin{document}

\title{A two-colour heterojunction unipolar nanowire light-emitting diode\\ by tunnel injection}
\author{Mariano A Zimmler, Jiming Bao, Ilan Shalish, Wei Yi, Venkatesh Narayanamurti, and Federico Capasso}\email{capasso@seas.harvard.edu}
\affiliation{Harvard University, Cambridge, MA 02138}

\date{\today}

\begin{abstract}

We present a systematic study of the current-voltage characteristics and electroluminescence of gallium nitride (GaN) nanowire on silicon (Si) substrate heterostructures where \emph{both} semiconductors are n-type. A novel feature of this device is that by reversing the polarity of the applied voltage the luminescence can be selectively obtained from either the nanowire or the substrate. For one polarity of the applied voltage, ultraviolet (and visible) light is generated in the GaN nanowire, while for the opposite polarity infrared light is emitted from the Si substrate. We propose a model, which explains the key features of the data, based on electron tunnelling from the valence band of one semiconductor into the conduction band of the other semiconductor. For example, for one polarity of the applied voltage, given a sufficient potential energy difference between the two semiconductors, electrons can tunnel from the valence band of GaN into the Si conduction band. This process results in the creation of holes in GaN, which can recombine with conduction band electrons generating GaN band-to-band luminescence. A similar process applies under the opposite polarity for Si light emission. This device structure affords an additional experimental handle to the study of electroluminescence in single nanowires and, furthermore, could be used as a novel approach to two-colour light-emitting devices.

\end{abstract}

\maketitle

\section{Introduction}

The generation of electroluminescence (EL) in semiconductor nanowires (NWs) has generally relied on p-n junction light-emitting diodes (LEDs) of different kinds. Several geometries and materials have been used, such as (\emph{i}) single-nanowire p-n junctions with doping modulation along the axial direction \cite{Haraguchi1992,Kim2003}, (\emph{ii}) crossed-wire junctions, where a p-type NW is placed across an n-type NW (both NWs not necessarily of the same semiconductor material) \cite{Huang2005}, (\emph{iii}) core-shell heterostructures with p-type and n-type layers \cite{Qian2005}, (\emph{iv}) structures consisting of an n-type NW in contact with a p-type substrate \cite{Duan2003, Bao2006, Zimmler2007, Motayed2007}. A common difficulty with some of these approaches, notably with (\emph{i}), is that it is not always possible to obtain both types of conductivity in the same semiconductor. Furthermore, in geometries such as (\emph{ii}) carrier injection is typically limited to a small fraction of the semiconductor material, greatly limiting output power and leading to poorly reproducible device characteristics. As we have previously shown \cite{Bao2006}, a n-type NW on a p-type substrate generally provides a feasible alternative that surmounts these problems. In this paper, we show it is possible to generate EL if also the substrate is of n-type conductivity, \emph{i.e. a unipolar LED}. A novel feature of this device is that by reversing the polarity of the applied voltage the luminescence can be selectively obtained from either component semiconductor. In this work, we use a gallium nitride (GaN) NW and a silicon (Si) substrate, so that for one polarity of the applied voltage ultraviolet (UV) (and visible) light is generated in the NW, while for the opposite polarity infrared (IR) light is emitted from the substrate.

As we show below, this device can be understood within the framework of a semiconductor-insulator-semiconductor (SIS) heterojunction. SIS diodes have received limited attention compared to p-n junctions and metal-insulator-semiconductor structures, with most studies concerned with their transport \cite{Shewchun1972a, Shewchun1972b, Shewchun1972c} and photovoltaic properties \cite{Shewchun1978, Shewchun1979}. To the best of our knowledge, no previous work has considered the possibility of using this unipolar structure for EL applications.

\section{Fabrication procedures}

\begin{figure}[bp]
   \centering
   \includegraphics[width=0.3\textwidth]{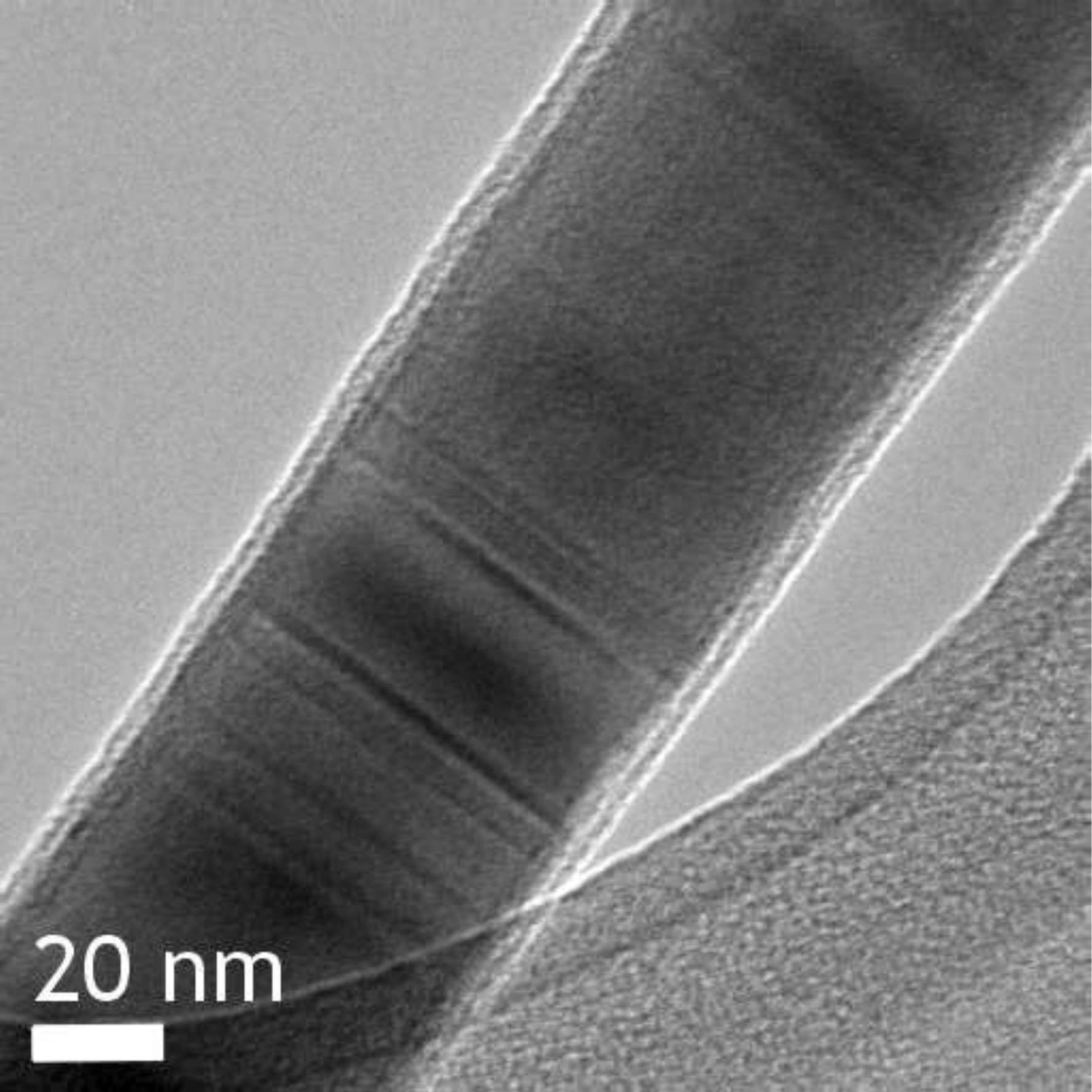}
   \caption{High-resolution TEM image of a GaN nanowire exhibiting a $\sim$5 nm shell of oxide.}
   \label{fig:tem}
\end{figure}

GaN NWs were grown by hydride vapour phase epitaxy (HVPE), unintentionally n-type doped with carrier concentrations in excess of 10$^{18}$ cm$^{-3}$ \cite{Seryogin2005}. Transmission electron microscope (TEM) images have revealed that a thin amorphous layer of oxide typically surrounds the NWs (figure \ref{fig:tem}). However, careful TEM characterization has also shown that the coverage of the oxide on the NWs is generally not uniform, and in some particular cases (for very thin NWs) it is even absent. This oxide, in fact, has measurable consequences for the behaviour of the device.

\begin{figure}[tbp]
   \centering
   \includegraphics[width=0.45\textwidth]{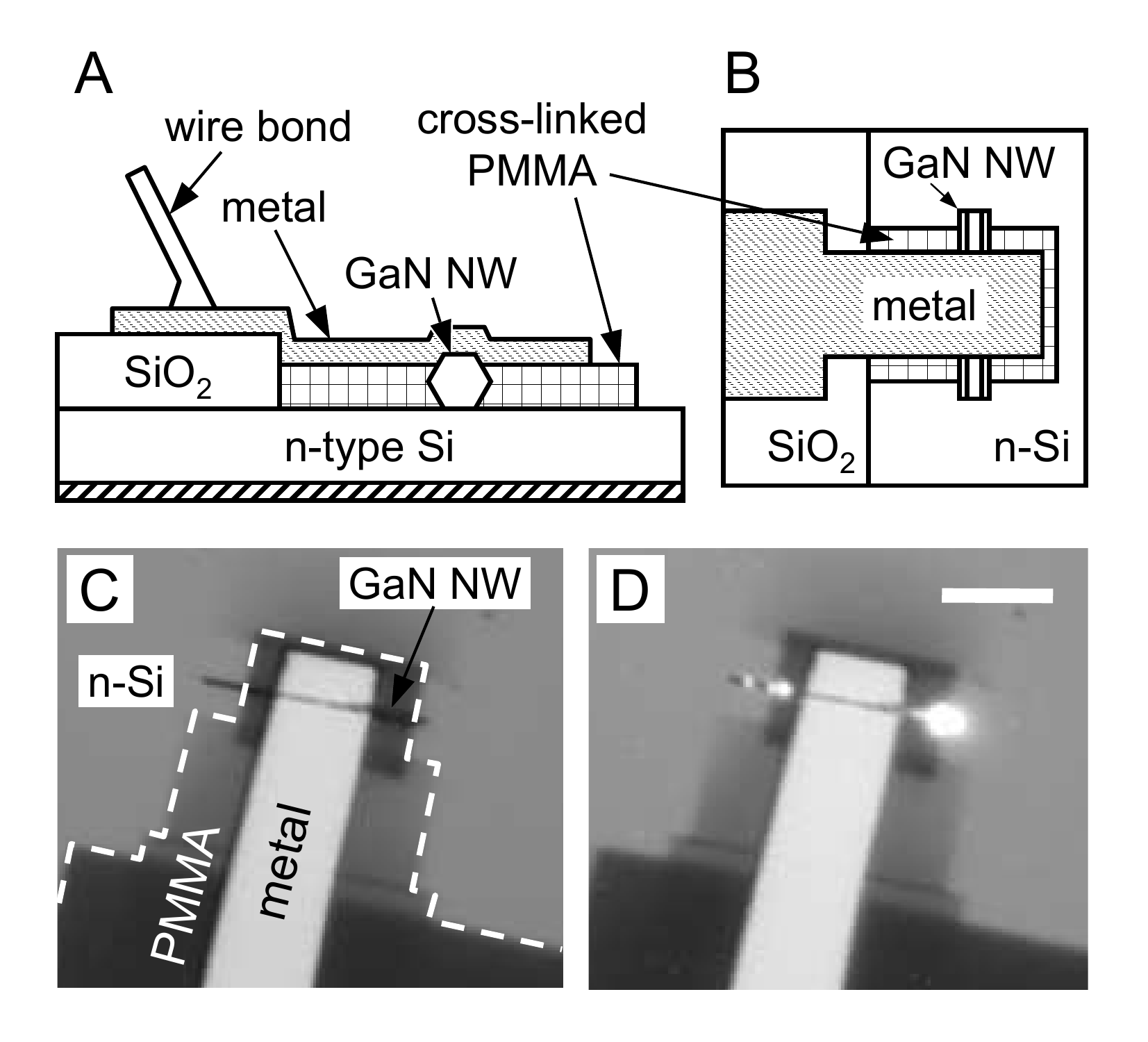}
   \caption{(A) Schematic cross-section and (B) top view of a completed device. (C) and (D) are optical microscope images of a device with no applied bias and under a positive bias, respectively. The scale bar is 5 $\mu$m. Note in (D) the light emission from the nanowire.}
   \label{fig:cs}
\end{figure}

Our device structure is based on a sandwich geometry in which a NW is placed between the substrate and a metallic contact, using cross-linked poly(methyl methacrylate) (PMMA) as an insulating spacer layer to prevent the metal contact from shorting to the substrate \cite{Bao2006}, as shown in figures \ref{fig:cs}A and B. This allows for uniform injection of current along the length of the NW.

Prior to transferring the NWs from the growth substrate onto the Si substrate for device assembly, a number of preparatory processing steps need to be carried out. We start with a heavily doped (10$^{19}$ cm$^{-3}$) n-type silicon substrate (n-Si) covered by a 200 nm thermal oxide. Photolithographic and etch steps (using buffered hydrofluoric acid) are used to create a square opening $\sim800 \times 800$ $\mu$m$^2$ through the thermal oxide, thereby exposing the underlying silicon. Subsequently, a pattern of markers is defined by means of electron-beam (e-beam) lithography and metal deposition in order to facilitate NW location after the NWs are transferred to the substrate, as well as to serve as alignment marks for later device processing. The NWs are then removed from the growth substrate using ultrasonic agitation in ethanol. The NW suspension thus obtained is applied dropwise to the processed wafer, resulting, after evaporation of the solvent in ambient air, in a random dispersion of NWs. This process brings the NWs into contact with the n-Si substrate via van der Waals forces and thus defines the bottom electrical contact to the NW.

The fabrication of the top contact to the NW relies on the use of PMMA as a negative (e-beam) resist and electrically insulating layer \cite{Bao2006}. Once the NWs have been dispersed onto the substrate, a thin layer of PMMA is spun on. The key step involves patterning the PMMA in a manner such that it is cross-linked on both sides of the NW but not on top of it (cross-linking PMMA can be achieved with exposure doses $\sim$10 mC/cm$^2$). After e-beam exposure the sample is immersed in acetone, which removes any PMMA that is not cross-linked (that is, on top of the NW). The preceding steps result in electrically insulating pads on both sides of the NW, which makes it possible to deposit a top metallic contact (Ti/Au) without it shorting to the substrate (see figures \ref{fig:cs}A and B). Device assembly is finalized by gluing the n-Si wafer back side onto a metallic sample holder with conducting silver paste, which also provides an ohmic contact the substrate. Figure \ref{fig:cs}A shows a cross-sectional view of the complete structure and figure \ref{fig:cs}B shows a corresponding top view. Figures \ref{fig:cs}C and D show optical micrographs of a typical device with no applied bias and under a positive bias voltage, respectively. Note that the latter corresponds to a positive potential of the n-Si substrate with respect to the top metallic contact, so that electrons flow from the top metallic contact into the NW and the substrate.

The EL measurement setup consisted of a 36X reflective microscope objective with an aluminum coating, coupled to a 1/4 m spectrometer (150 lines mm$^{-1}$ grating) and a thermoelectrically cooled CCD camera. The images in the insets of figure \ref{fig:iv} were obtained utilizing the zeroth-order beam of the grating, whereas the device emission spectra in figure \ref{fig:ccdspectra} were obtained utilizing the first-order beam. The spectra in figure \ref{fig:igaspectra} were recorded with an InGaAs detector ($\sim$0.7 A/W responsivity at 1100 nm) in combination with a lock-in amplifier. Low-temperature measurements were carried out in a continuous flow cryostat with optical access to the sample.

\section{Results}

Five devices were fabricated as described above and characterized in detail by means of current-voltage characteristics and electroluminescence. The behavior observed in these devices could be categorized into two classes, which we shall refer to henceforth as ``type I" and ``type II". As we discuss further on, the differences between the two classes may be associated with the presence or absence of an interfacial oxide under the metallic top contact, due to a non-ideal fabrication process.\footnote{Note: PMMA is susceptible to most acids and is not stable at high temperatures. Therefore, the standard steps of oxide etching and thermal annealing commonly involved in contact metallization had to be avoided in our case. Alternative insulators are currently under investigation.} Nevertheless, we believe it is important for the completeness of the analysis to present and compare both cases, as both may still be relevant in the fabrication of today's nanowire devices, which is presently at its infancy. As a matter of consistency, however, all data presented here correspond only to two devices, one of each type.

\subsection{Current-voltage characteristics}

\begin{figure}[tbp]
   \centering
   \includegraphics[width=0.45\textwidth]{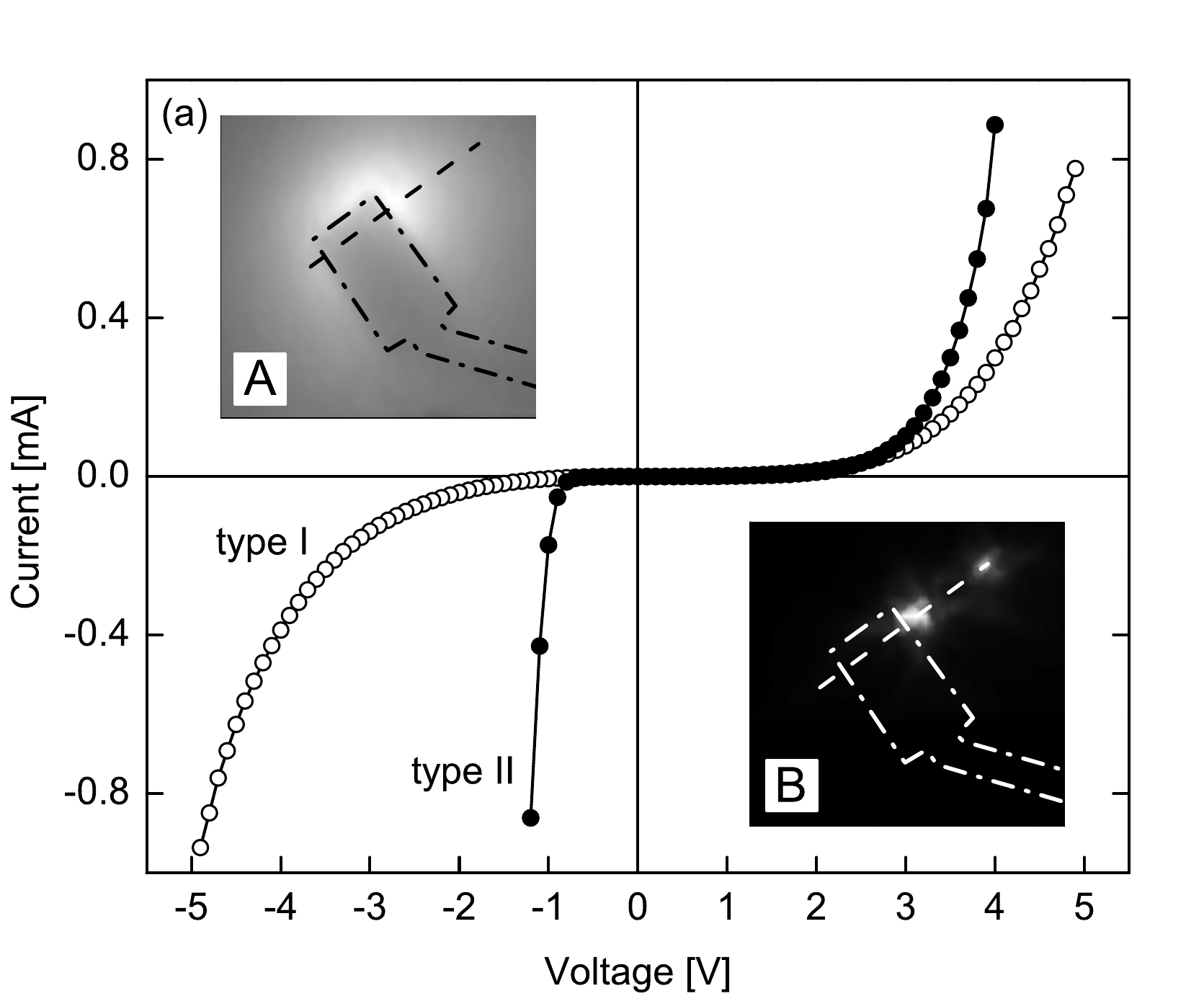}
   \includegraphics[width=0.45\textwidth]{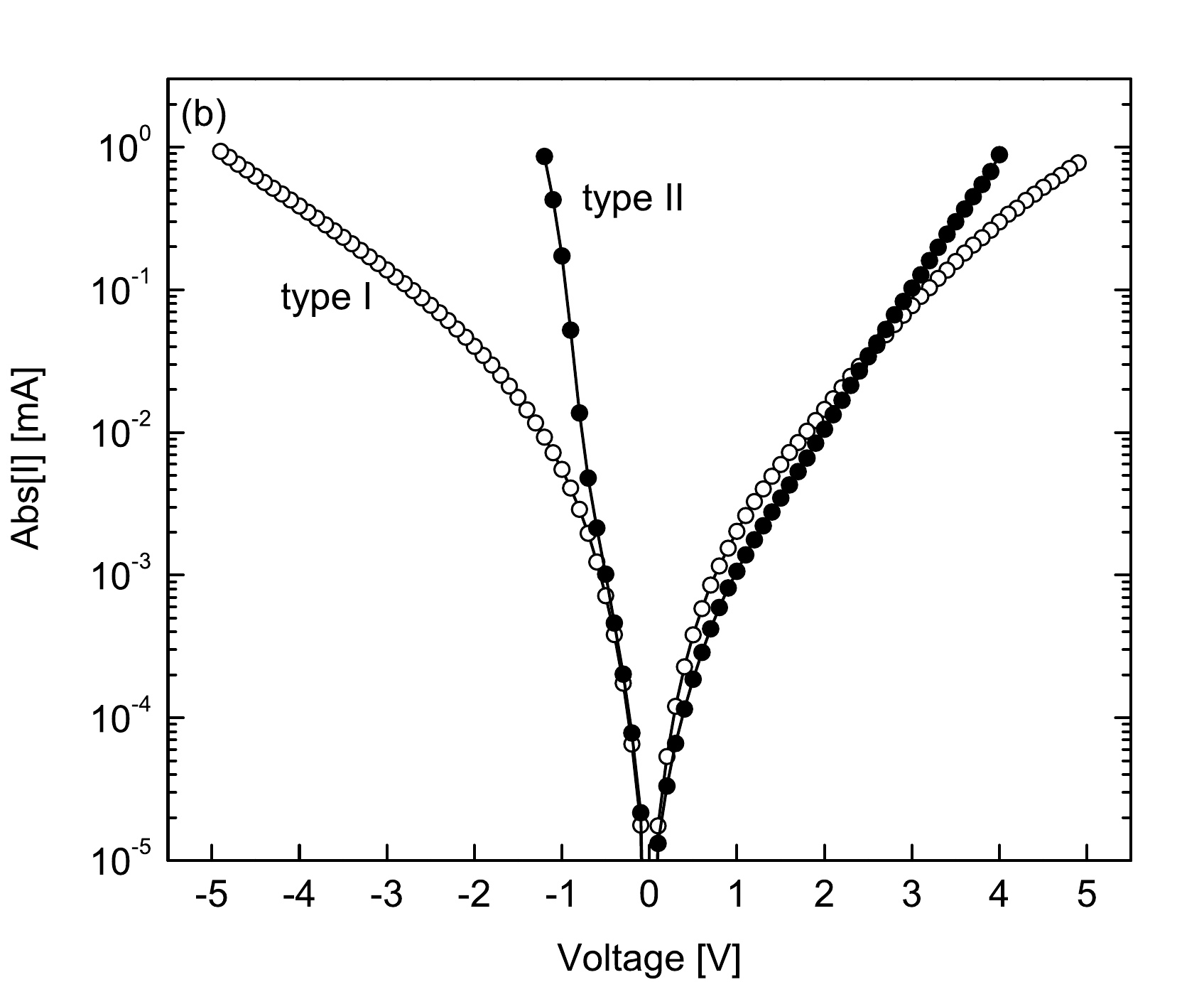}
   \caption{Current-voltage characteristics of two representative n-GaN NW/n-Si substrate devices plotted in (a) linear and (b) semi-logarithmic scale. The insets in (a) are optical micrographs collected by a CCD camera of a type II device under (A) negative and (B) positive values of the applied voltage.}
   \label{fig:iv}
\end{figure}

Room-temperature current-voltage (I-V) characteristics of two representative devices are shown in figure \ref{fig:iv}. Under moderate positive applied voltage ($V<2E_{\mathrm{g}}^{\mathrm{GaN}}/e$, where $E_{\mathrm{g}}^{\mathrm{GaN}} \sim 3.4$ eV is the bandgap of GaN at room temperature \cite{Madelung2003} and $e$ is the electron charge) the current increases approximately exponentially with voltage for both types of devices. When the applied voltage exceeds $\sim$3 V, both types of devices exhibit EL originating from the NW. For example, inset B in figure \ref{fig:iv} shows an optical micrograph collected by a CCD camera of a type II device, taken at room temperature and under an applied voltage of $+4$ V. Light emission from the NW (indicated by a dashed line) occurs at two locations: from near the metallic contact (defined by the dash-dot line) and from the right end of the NW. Under the opposite voltage polarity, the I-V characteristics differ markedly for both types of devices. In the case of type I devices, the response is similar to the positive voltage behaviour, whereas type II devices exhibit a sharp increase in the magnitude of the current with an applied voltage of about $-1$ V. Despite this difference, both types of devices emit light from a broad, diffuse area in the substrate: type II devices when the applied voltage is $< -1$ V and type I devices when  $V < -2$ V. Inset A in figure \ref{fig:iv} is an image of a type II device under an applied voltage of $-1.3$ V, at room temperature. Current-voltage characteristics obtained at lower temperatures (77 K and 7 K) exhibited nearly unchanged current levels, revealing that conduction through either type of device is not thermally activated.

\subsection{Electroluminescence spectra: positive and negative voltage}

\begin{figure}[tbp]
   \centering
   \includegraphics[width=0.45\textwidth]{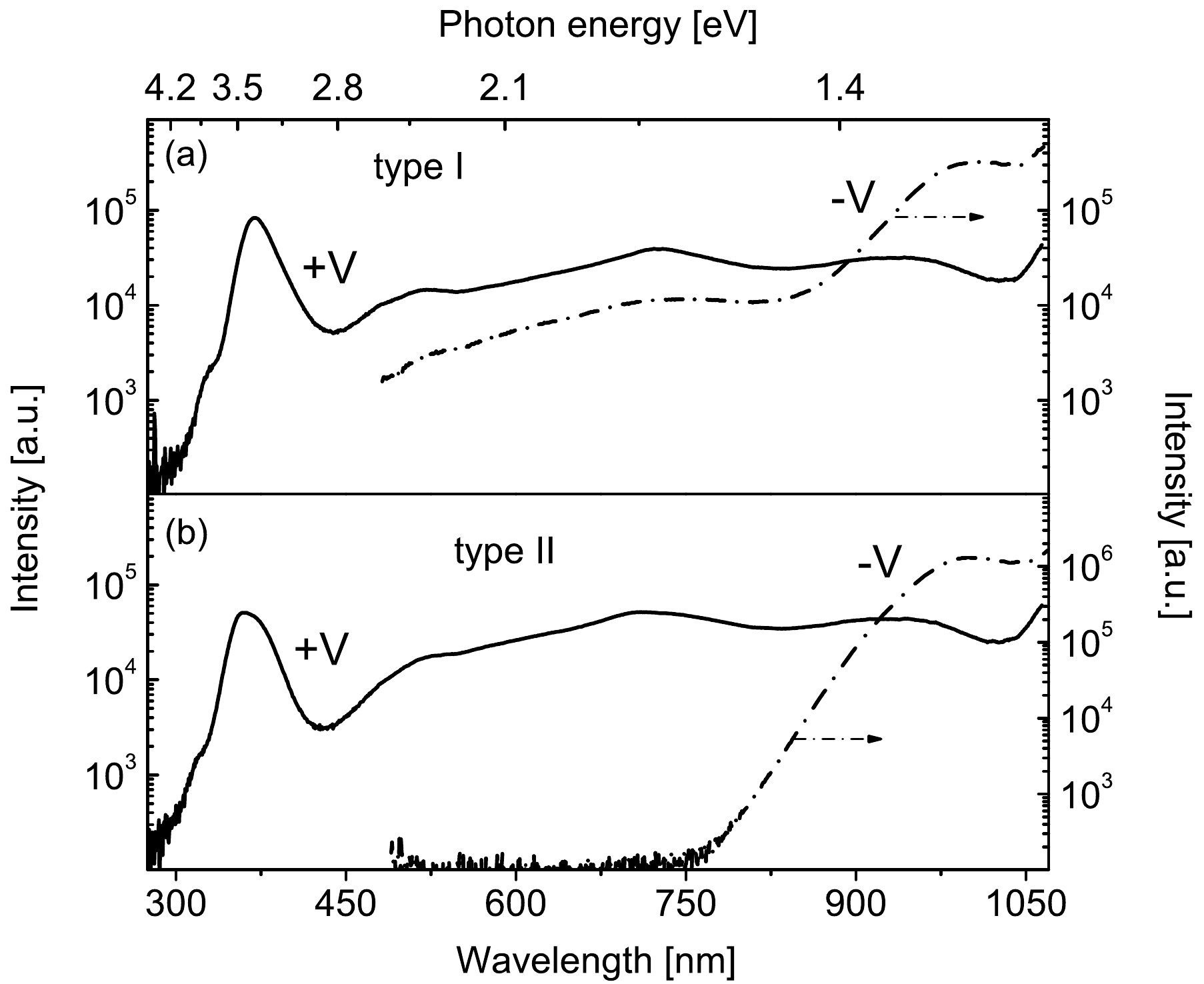}
   \caption{Electroluminescence (EL) spectra of type I and II devices at room temperature, obtained with a Si-CCD array. (a) EL spectra of a type I device for $V=+5.2$ V, $I=900$ $\mu$A (solid line) and $V=-4.8$ V, $I=-960$ $\mu$A (dash-dot line). (b) EL spectra of a type II device for $V=+4.2$ V, $I=820$ $\mu$A (solid line) and $V=-1.15$ V, $I=-936$ $\mu$A (dash-dot line). EL spectra obtained under different applied voltages were similar in shape. All measured intensities have been corrected for the CCD and grating spectral response.}
   \label{fig:ccdspectra}
\end{figure}

Figure \ref{fig:ccdspectra} shows the recorded EL spectra for type I and II devices, at room temperature, under positive and negative applied voltages. The spectral characteristics of the light emission under positive voltage (solid lines) are very similar for both types of devices: the spectra consist of a UV peak centred at $\sim$360 nm and a broad-band feature extending from $\sim$500 nm to about 950 nm, with a maximum around 720 nm. The UV luminescence is consistent with band-to-band recombination in GaN. The broad sub-bandgap luminescence is most likely due to radiative recombination between deep levels, similar to what is commonly observed in the photoluminescence of HVPE GaN \cite{Reuter1999, Gotz1996, Rhee1998}. We estimate the room-temperature external quantum efficiency (defined as the number of photons emitted into free space per second divided by the number of electrons injected per second) \emph{in the UV range} to be about 10$^{-8}$ for an applied voltage of $\sim$5 V. The light collection efficiency of the EL measurement setup is $\sim$1$\%$.

\begin{figure}[tbp]
   \centering
   \includegraphics[width=0.45\textwidth]{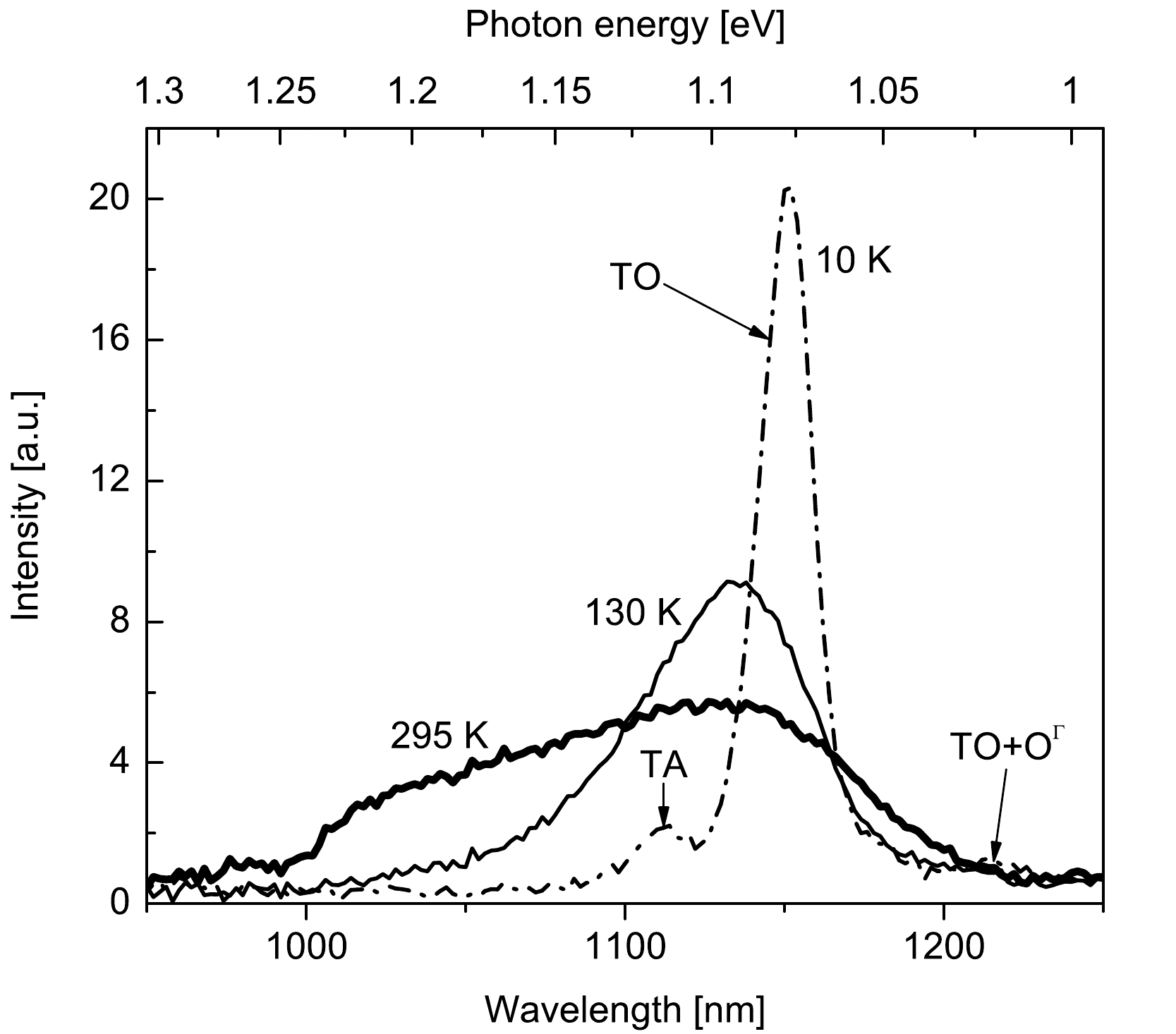}
   \caption{Negative voltage electroluminescence spectra of a type II device, at several temperatures, obtained with an InGaAs detector. The measured intensities have been corrected for the InGaAs detector spectral response.}
   \label{fig:igaspectra}
\end{figure}

The spectral characteristics of the light emission for negative values of the applied voltage (dash-dot lines) is, in contrast, somewhat different for type I and II devices. Clearly, both types of devices exhibit significant EL at wavelengths in excess of $\lambda=900$ nm. However, type I devices also show a feature, \emph{similar to the positive voltage broad band emission} (centred at $\sim$720 nm), that is absent in the spectral characteristics of type II devices. No GaN bandgap luminescence was detected, however, in either type. Figure \ref{fig:ccdspectra} therefore suggests that, in the case of type I devices, some of the light emission mechanisms at work under positive voltage are also present under negative voltage. To make a more accurate identification of the IR luminescence, we have repeated the measurement of the negative voltage luminescence using an InGaAs detector, since the quantum efficiency of the Si-CCD array drops below $0.1\%$ beyond 1080 nm. Figure \ref{fig:igaspectra} shows the EL spectra of a type II device at various temperatures, all obtained with a fixed current of 1 mA. The 10 K spectrum reveals features commonly observed in silicon luminescence experiments: the two-phonon transverse optic and zone center optic (TO+O$^{\Gamma}$), transverse optic (TO), and transverse acoustic (TA) phonon assisted transitions at photon energies $\sim$1.02 eV, $\sim$1.08 eV, and $\sim$1.12 eV, respectively \cite{Davies1989}. This confirms that the light emission under negative voltage results from radiative recombination \emph{in the substrate} as was inferred from inset A in figure \ref{fig:iv}. The estimated room-temperature external quantum efficiency in the IR range (at $-1.34$ V) is about 10$^{-6}$.

\begin{figure}[tbp]
   \centering
   \includegraphics[width=0.45\textwidth]{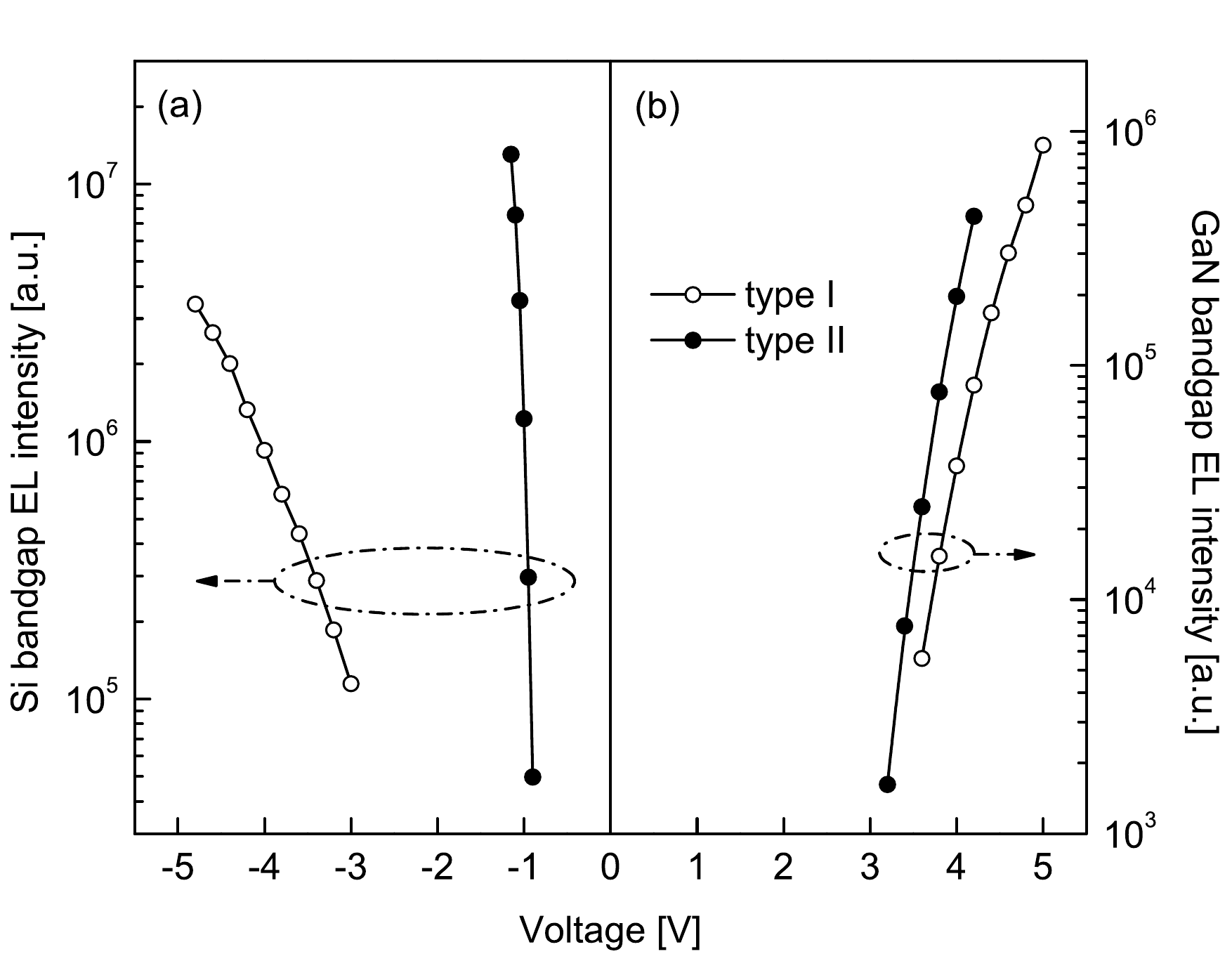}
   \caption{Integrated EL intensity, at room temperature, as a function of voltage for type I and II devices. The intensities were obtained by integrating the counts measured with the Si-CCD array: in the range from 300 nm to 450 nm for positive voltages (\emph{i.e.} GaN bandgap luminescence) and from 800 nm to 1150 nm for negative voltages (\emph{i.e.} Si bandgap luminescence).}
   \label{fig:lv}
\end{figure}

Figure \ref{fig:lv} shows the voltage dependence of band-to-band EL for positive bias (\emph{i.e.} GaN band-to-band EL) and negative voltage (\emph{i.e.} Si band-to-band EL). Two aspects of this figure are important to note: first, for the type II device EL turns on very abruptly at an applied voltage of about $-1$ V; second, a type I device requires a larger applied voltage to achieve similar output intensities: a fraction of a volt for positive voltage and several volts for negative applied voltage.

\begin{figure*}[tbp]
   \centering
   \includegraphics[width=0.85\textwidth]{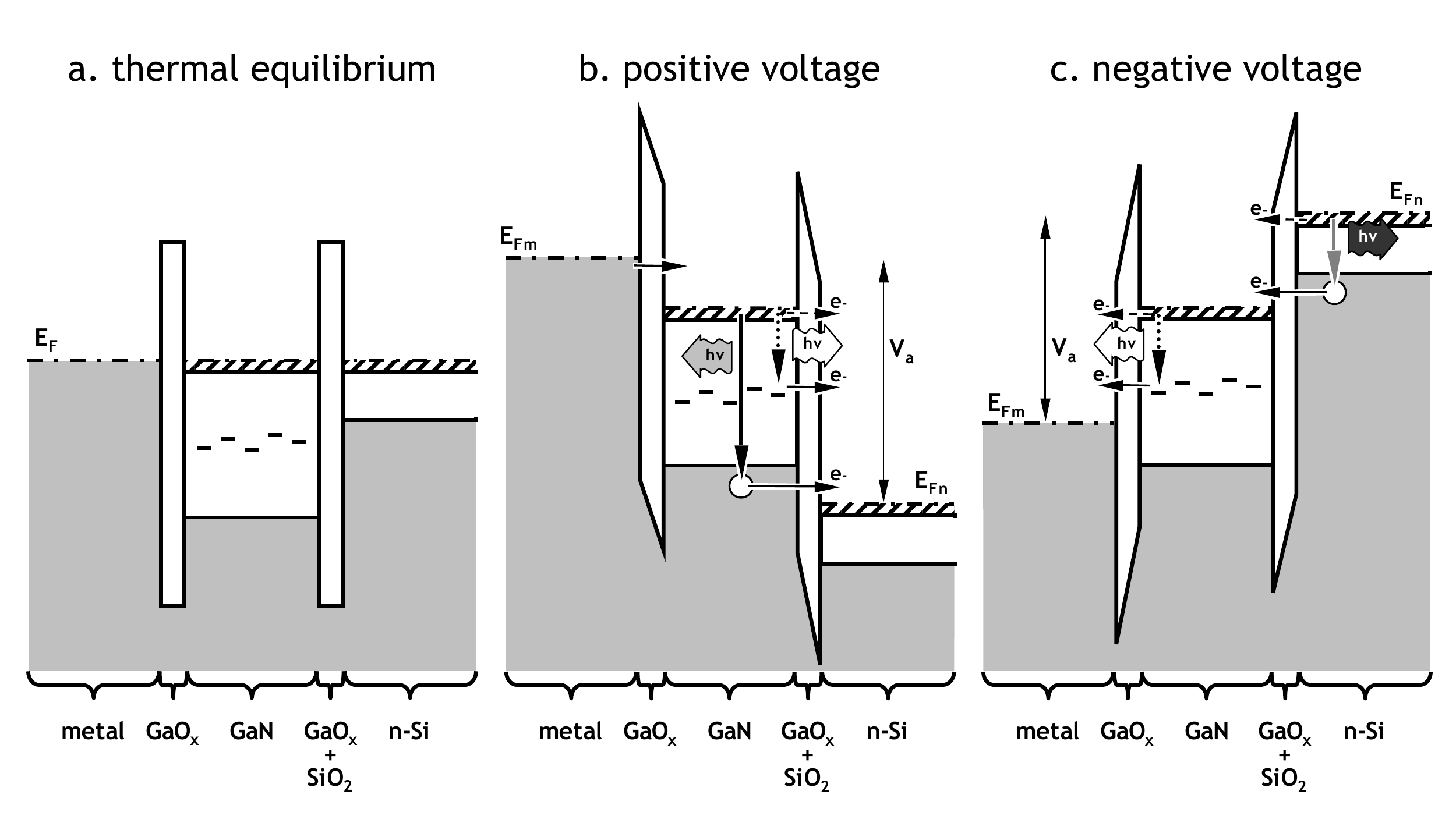}
   \caption{Schematic band diagram of a n-GaN/n-Si unipolar electroluminescent device of type I (a) in thermal equilibrium, (b) under positive applied voltage and (c) under negative applied voltage. In (b) the solid black downward arrow indicates band-to-band radiative transitions in GaN. The dotted black downward arrows in (b) and (c) indicate deep-level radiative transitions in GaN. The solid grey downward arrow in (c) indicates band-to-band radiative transitions in Si. Horizontal arrows indicate tunnelling of electrons.}
   \label{fig:bd1}
\end{figure*}

\section{Discussion}

The interpretation of these observations requires the realization that junctions that result from bringing two semiconductors into contact (the n-type NW and the n-type substrate, in our case) are far from ideal compared with junctions epitaxially grown in planar systems. In contrast to the latter, covalent chemical bonds between the two semiconductors are not formed. Rather, the interface region arises from a mechanical contact via van der Waals forces. In addition, upon exposure to ambient air, surfaces of semiconductors normally develop thin layers of native oxide ($\sim$ 1-2 nm of SiO$_2$ on Si and a few nm of GaO$_x$ surrounding GaN NWs, as evidenced in figure \ref{fig:tem} and previous work \cite{Watkins1999, Tang2003}), which transform the system into an effective SIS structure. We should emphasize that such oxides are unavoidably present both on NWs and on substrates, so that systems relying on \emph{ex situ} assembly of a device from two such components will typically behave consistently with an SIS system (the same is true, for example, of crossed-wire junctions). This aspect has so far been largely neglected in previous studies of these systems\cite{Huang2005,Duan2003}.

\subsection{Band diagrams for type I devices}
\label{sec:bd1}

Figure \ref{fig:bd1} shows our proposed simplified band diagrams for a n-GaN/n-Si device along the NW radial direction perpendicular to the substrate, which explain the key features of the data. Thin oxide barriers both between the NW and the substrate and between the NW and the metal are included. Band bending, image forces and potential energy drops across the oxide layers in equilibrium are neglected for simplicity. Figure \ref{fig:bd1}(a) is a sketch of the thermal equilibrium band diagram of the device. Under an applied bias, most of the voltage drops across the oxide layers since both the Si and the GaN are heavily doped and therefore highly conductive.

For positive values of the applied voltage (figure \ref{fig:bd1}(b)), due to the interface oxide between the NW and the Si substrate, the conduction band of Si can be substantially lowered below that of GaN. In particular, when the potential energy drop between the two semiconductors exceeds $\sim$3.4 eV (\emph{i.e.} the band gap energy of GaN), the Si quasi-Fermi energy is lowered below the valence band edge in GaN. This makes it possible for valence band electrons in GaN to tunnel into the conduction band of Si, which is equivalent to hole back-injection into GaN. The injected holes can then recombine radiatively with conduction band electrons generating band-to-band luminescence. This process is indicated schematically in figure \ref{fig:bd1}(b) by the solid black downward arrow. For the type I device in figures \ref{fig:ccdspectra}(a) and \ref{fig:lv}(b), the threshold voltage for GaN band-to-band EL is $\sim$+3.5 V.

The broad band sub-bandgap emission exhibited by type I devices for positive voltage (solid line in figure \ref{fig:ccdspectra}(a)) can be explained in a similar manner. As the Si quasi-Fermi level traverses the GaN bandgap with increasing applied voltage, increasingly many electrons trapped in deep levels near the surface of GaN can tunnel into the conduction band of Si, leaving behind holes which can recombine radiatively with conduction band electrons (indicated schematically by the dotted black downward arrow in figure \ref{fig:bd1}(b)).\footnote{In our previous paper \cite{Zimmler2007}, we argued that the mechanism for the generation of broad sub-bandgap emission in n-type GaN NW on p-type Si substrate LEDs involved the recombination of electrons in deep levels near the surface with valence band holes. However, it is more likely that a mechanism as described here is responsible for it.}

For negative values of the applied voltage (figure \ref{fig:bd1}(c)) the situation reverses. In this case, the conduction band of GaN is lowered with respect to that of Si. When the potential energy drop between the two semiconductors exceeds $\sim$1 eV (\emph{i.e.} the band gap energy of Si), the GaN quasi-Fermi energy is lowered below the valence band edge of Si. Valence band electrons in Si can then tunnel into the conduction band of GaN, leaving behind holes that can recombine radiatively with conduction band electrons, generating band-to-band luminescence near the silicon bandgap (this is indicated schematically by the solid grey downward arrow in figure \ref{fig:bd1}(c)). The threshold voltage for Si band-to-band EL in the device in figures \ref{fig:ccdspectra}(a) and \ref{fig:lv}(a) is $\sim$-2 V. Note that this value is larger than the bandgap energy of Si divided by the electron charge. The reason for this is the presence of the additional interface oxide between the NW and the metal, which results in only a fraction of the applied voltage being across the NW/Si interface.

\begin{figure}[tbp]
   \centering
   \includegraphics[width=0.45\textwidth]{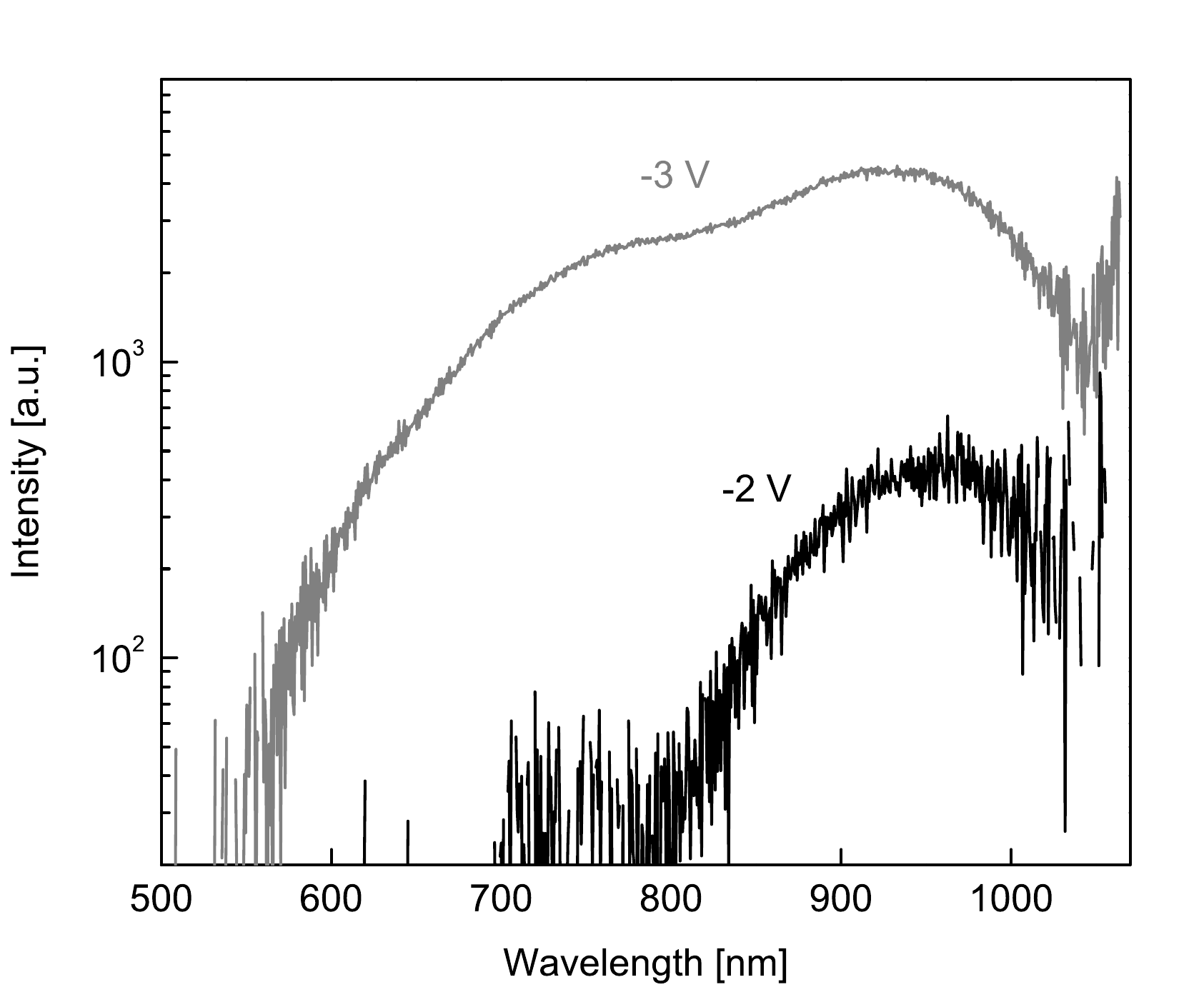}
   \caption{Electroluminescence spectra of the type I device at 7 K, for low applied negative voltages, recorded with a Si-CCD camera. The spectra have been corrected for the CCD and grating spectral response.}
   \label{fig:spectrav}
\end{figure}

As seen in figure \ref{fig:ccdspectra}(a), type I devices emit broad band light below the GaN bandgap also under negative voltage. This can be explained by the presence of the additional interface oxide between the NW and the metallic contact, which makes it possible to sustain a potential energy difference between the NW and metal quasi-Fermi levels. With increasing negative voltage, the metal Fermi level traverses the GaN bandgap, so that increasingly many electrons trapped in deep levels near the surface of GaN can tunnel into the metal. Once again, this results in GaN holes that can recombine radiatively with conduction band electrons (dotted black downward arrow in figure \ref{fig:bd1}(c)). Effectively, the NW/metal interface under negative voltage is equivalent to the NW/Si junction under positive voltage and therefore results in similar luminescence characteristics. Figure \ref{fig:spectrav} provides further reinforcement for this argument. For low values of the applied voltage ($V\sim-2$ V, $I\sim-79$ $\mu$A) the emission intensity first peaks at $\sim$915 nm, whereas, upon increasing the magnitude of the applied voltage further ($V\sim-3$ V, $I\sim-290$ $\mu$A) the peak at $\sim$720 nm appears and grows in intensity.

\begin{figure}[tbp]
   \centering
   \includegraphics[width=0.45\textwidth]{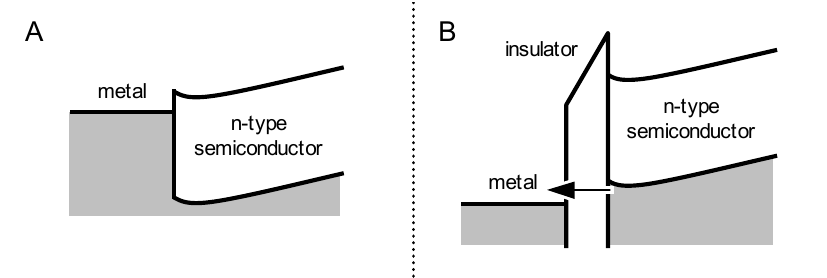}
   \caption{(A) Schematic band diagram of a metal-semiconductor junction with a forward voltage applied. (B) Band diagram of a metal-insulator-semiconductor junction with a forward voltage applied. The insulator makes it possible for valence band electrons to tunnel to the metal. This process is indicated by the arrow.}
   \label{fig:ms_mis}
\end{figure}

Similar phenomena have been observed in the past in planar metal-insulator-semiconductor diodes (so-called MIS structures). When a metallic contact is deposited directly onto a n-type semiconductor surface (figure \ref{fig:ms_mis}(A)), it is normally impossible for holes to be injected from the metal into the semiconductor by application of small voltages. The reason is that the holes would have to be thermally excited over a barrier whose height is equal to the bandgap energy minus the electron Schottky barrier height. However, Jaklevic \emph{et al.} \cite{Jaklevic1963} first realized that the inclusion of an insulating layer between the metal and the semiconductor (figure \ref{fig:ms_mis}(B)) makes it possible to maintain a sufficient potential energy difference so that electron tunnelling from the semiconductor valence band into the metal is energetically possible. In particular, when the voltage across the oxide is large enough to provide the band diagram in figure \ref{fig:ms_mis}B, such a process would take place, with the subsequent radiative recombination of holes and conduction band electrons. Several later works confirmed these results \cite{Fischer1963, Card1971, Rosenkrantz1968, Clark1976, Livingstone1973}.

The NW/metal interface in devices of type I is, in fact, exactly an MIS junction, due to the presence of the interface oxide. The NW/substrate interface can also be understood in the language of MIS structures: the GaN and the n-Si alternatively assume the roles of the semiconductor and the metal depending on the bias polarity. For positive voltage, the GaN assumes the role of the semiconductor and the n-Si acts as the metal. For negative voltage, the opposite is true: the n-Si assumes the role of the semiconductor and the GaN acts as the metal.

\subsection{Band diagrams for type II devices}
\label{sec:bd2}

\begin{figure*}[tbp]
   \centering
   \includegraphics[width=0.85\textwidth]{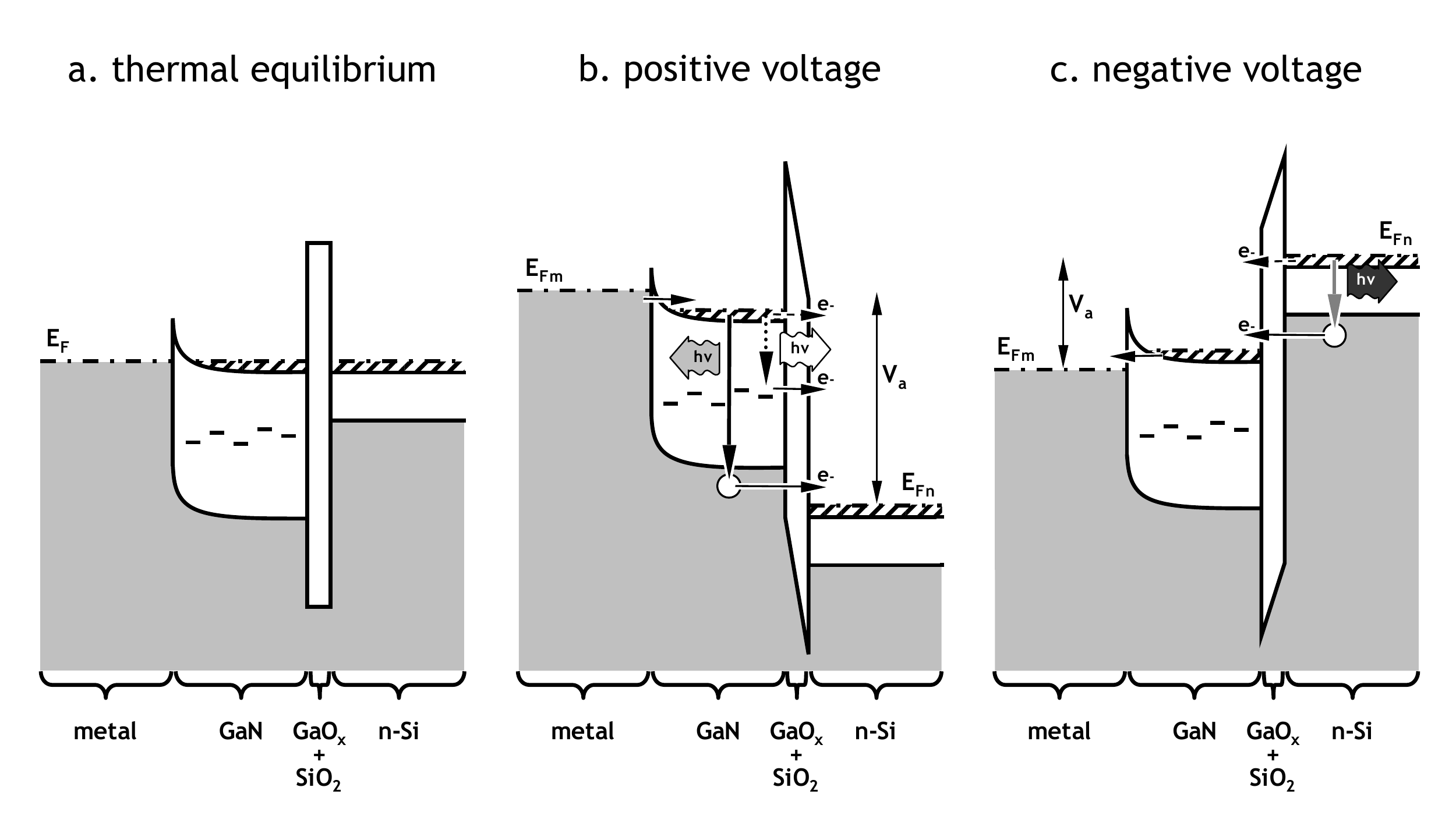}
   \caption{Schematic band diagram of a n-GaN/n-Si unipolar electroluminescent device of type II (a) in thermal equilibrium, (b) under positive voltage and (c) under negative voltage. In (b) the solid black downward arrow indicates band-to-band radiative transitions in GaN and the dotted black downward arrow indicates deep-level radiative transitions. The solid grey downward arrow in (c) indicates band-to-band radiative transitions in Si. Horizontal arrows indicate tunnelling of electrons.}
   \label{fig:bd2}
\end{figure*}

Before proceeding, let us summarize the properties of the two types of devices so far discussed: (\emph{i}) the current is considerably higher in type II devices than it is in type I devices (figure \ref{fig:iv}); (\emph{ii}) the electroluminescence spectrum from both types of devices (figure \ref{fig:ccdspectra}) is similar for positive applied voltages, with the only difference being that the voltage threshold for GaN band-to-band light emission is lower for type II devices (see figure \ref{fig:lv}); (\emph{iii}) devices of type II \emph{do not exhibit broad-band GaN sub-bandgap light for negative applied voltages} (figure \ref{fig:ccdspectra}), but both devices emit IR luminescence from the substrate. These points, as well as the previous discussion of metal-semiconductor and metal-insulator-semiconductor junctions, suggest that the main structural difference between the two types of devices resides at the NW/metal interface. In particular, these features strongly support the conclusion that the NW/metal interface oxide is absent in type II devices. Figure \ref{fig:bd2} depicts the corresponding band diagrams for a device lacking the NW/metal oxide, but otherwise identical to the device in figure \ref{fig:bd1}.

Under positive voltage (figure \ref{fig:bd2}(b)), type II devices behave in much the same way as type I devices (compare with figure \ref{fig:bd1}(b)), with the only exception being that the additional ohmic drop across the NW/metal oxide in type I devices is now absent. This should therefore result in a lower voltage threshold for band-to-band luminescence, which is indeed observed in figure \ref{fig:lv}. Band-to-band and deep level luminescence proceed in the same way as in devices of type I. The reader is referred for this description to section \ref{sec:bd1}.

Under negative voltage (figure \ref{fig:bd2}(c)) type II devices behave very differently from those of type I (compare with figure \ref{fig:bd1}(c)). The lack of an insulating layer between the NW and the metal makes it impossible (with the application of low voltages) for electrons to be injected from deep levels near the GaN surface into the metal so that no sub-bandgap luminescence is produced in the GaN. As for type I devices, under negative applied voltage the conduction band of GaN is lowered with respect to that of Si, due to the presence of the oxide in the NW/substrate interface. When the potential energy drop between the two semiconductors exceeds $\sim$1 eV (\emph{i.e.} the band gap energy of Si), the GaN quasi-Fermi energy is lowered below the valence band edge of Si giving place to electron tunnelling from the Si valence band. The holes created in this process can then recombine radiatively with conduction band electrons generating Si band-to-band luminescence, as indicated schematically by the solid grey downward arrow in figure \ref{fig:bd2}(c). The main difference is that essentially all the applied voltage drops across the NW/substrate interface, so that the onset of Si band-to-band luminescence should be very close to the bandgap of Si divided by the electron charge. As evidenced by figure \ref{fig:lv}, the threshold for Si band-to-band luminescence is $\sim-$1 V. In this configuration, the system is equivalent to the well studied metal-oxide-silicon tunnelling diodes, in which silicon EL has been observed with similar properties as discussed here \cite{Chen2001, Chen2003}.

\subsection{General comments}

As we have described in sections \ref{sec:bd1} and \ref{sec:bd2}, the NW/substrate interface (regardless of the type of device) is responsible for the generation of band-to-band luminescence in both the GaN NW and the Si substrate. Under positive voltage, GaN band-to-band luminescence is produced as a consequence of electrons tunnelling from the GaN valence band into the Si conduction band. Similarly, under negative voltage, the tunnelling of electrons from the Si valence band into the GaN conduction band leads to the generation of Si band-to-band light. It is possible to obtain further insight into these tunnelling processes by analyzing the quantum efficiency of the device under different values of the applied voltage. For any LED, the external quantum efficiency can be defined as \cite{Bhattacharya1997}
\begin{equation}
\eta_{\mathrm{ext}} = \frac{W/(h\nu)}{I/e} = \eta_{\mathrm{int}} \eta_{\mathrm{extrac}},
\end{equation}
where $W$ is the optical power emitted into free space, $h\nu$ is the energy of a photon and $I$ is the current. The measured optical power $W_{\mathrm{meas}}$ is related to $W$ by the collection efficiency $\eta_{\mathrm{coll}}$, $W_{\mathrm{meas}} = \eta_{\mathrm{coll}}W$. The extraction efficiency $\eta_{\mathrm{extrac}}$ is given by the ratio $W/W_{\mathrm{int}}$, with $W_{\mathrm{int}}$ the optical power generated in the active region. The internal quantum efficiency $\eta_{\mathrm{int}}$ is given by the product of the injection efficiency $\eta_{\mathrm{injec}}$ and the radiative recombination efficiency $\eta_{\mathrm{rad}}$, that is, $\eta_{\mathrm{int}} = \eta_{\mathrm{injec}} \eta_{\mathrm{rad}}$ \cite{Bhattacharya1997}. Thus, under the reasonable assumption that $\eta_{\mathrm{extrac}}$ and $\eta_{\mathrm{rad}}$ are approximately constant with applied voltage, it follows that
\begin{equation}
\frac{W_{\mathrm{meas}}}{I} \propto \eta_{\mathrm{injec}}.
\label{eq:Wppetainj}
\end{equation}
For the NW/substrate junction, it is clear that the injection efficiency is given by the fraction of the total current that is carried by holes. With equivalent language, this current is given by electrons tunnelling from the GaN valence band into Si, under positive voltage, and by electrons tunnelling from the Si valence band into the NW, under negative voltage. If we denote either current by $I_{v \to c}$, we have for a given voltage polarity,
\begin{equation}
\eta_{\mathrm{injec}} = \frac{I_{v \to c}}{I_{v \to c} + I_{c \to c}}.
\label{eq:Iinj}
\end{equation}
The other component of the current $I_{c \to c}$ is given by the electrons tunnelling from the conduction band of one semiconductor into the conduction band of the other semiconductor. This process is indicated schematically by dashed horizontal arrows in figures \ref{fig:bd1}(b) and (c), and \ref{fig:bd2}(b) and (c). Evidently, the barrier for such tunnelling electrons is significantly smaller than the barrier for electrons tunnelling from the valence band, so that the injection efficiency for this kind of structure is bound to be much smaller than unity, leading to the very small measured external quantum efficiency. Finally, combining (\ref{eq:Wppetainj}) and (\ref{eq:Iinj}) we obtain
\begin{equation}
W_{\mathrm{meas}} \propto I_{v \to c}.
\end{equation}
$I_{v \to c}$ can be described as a Fowler-Nordheim type tunnelling current \cite{Zimmler2007} and therefore will exhibit an exponential dependence with applied voltage. Figure \ref{fig:lv} is a plot of $W_{\mathrm{meas}}$ as a function of $V$ which shows this. Note also that the curves are concave downwards, which is characteristic of a tunnelling process.

A final comment on device stability is in order. Even though most devices exhibited stable I-V characteristics, some of them did show sudden changes in conduction during operation. Typically, these changes in conduction were characterized by an increase or decrease of the voltage \emph{for a fixed current level}. More importantly, the increase or decrease of the voltage was also accompanied by a corresponding increase or decrease, respectively, of luminescence. This is also consistent with the framework that we have developed: electroluminescence intensity depends crucially on the ability to maintain a potential energy difference between the two semiconductors. When a change in conduction occurs, it implies that the current now flows through a different pathway with, for example, a slightly thicker oxide, which drops a larger potential for the same current level and therefore facilitates hole injection and subsequent radiative recombination.

\section{Conclusions}

We have presented a detailed analysis of a structure based on a n-type GaN NW on a n-type Si substrate. This structure emits UV (and visible) light from the GaN NW under one polarity of the applied voltage and IR light from the Si substrate under the opposite polarity. The key features of the data can be explained with a model based on electron tunnelling from the valence band of one semiconductor into the conduction band of the other semiconductor. For one polarity of the applied voltage, electrons in the GaN valence band can tunnel into the Si conduction band when the potential energy difference between the GaN and Si quasi-Fermi levels is approximately equal to the GaN bandgap energy. Those tunnelling electrons leave behind holes that can recombine with conduction band electrons generating GaN band-to-band light. Similarly, for the opposite voltage polarity, electrons in the Si valence band can tunnel into the GaN conduction band when the potential energy difference between the respective quasi-Fermi levels is approximately equal to the Si bandgap energy. This process also results in the creation of holes in the Si valence band that can recombine radiatively with conduction band electrons. The quantum efficiency of such a device is limited to be much smaller than unity due to the larger tunnelling barrier for valence band electrons as compared to conduction band electrons.

\begin{acknowledgments}

This work was supported by the National Science Foundation under grant no. ECS-0322720 and by the National Science Foundation Nanoscale Science and Engineering Center (NSEC) under contract NSF/PHY 06-46094. The support of the Center for Nanoscale Systems (CNS) at Harvard University is also gratefully acknowledged. Harvard-CNS is a member of the National Nanotechnology Infrastructure Network (NNIN).

\end{acknowledgments}

\end{document}